# *Molecular Dynamics of Spin Crossover:*
# *the (P,T) phase diagram of [Fe(PM-BIA)$_2$(NCS)$_2$].*

A. Marbeuf[a], P. Négrier[a], S.F. Matar[b,c], L. Kabalan[b,c], J.F. Létard[b,c], P. Guionneau[b,c]

[a]Université de Bordeaux-CNRS, LOMA, 351 cours de la Libération F-33400 Talence, France
[b] CNRS, ICMCB, UPR 9048, F-33600 Pessac, France
[c] Univ. Bordeaux, ICMCB, UPR 9048, F-33600 Pessac, France

Corresponding author: matar@icmcb-bordeaux.cnrs.fr

**Abstract**

*The spin crossover properties and the domains of existence of the different phases for the [Fe(PM-BIA)$_2$(NCS)$_2$] complex are obtained from combining DFT and classical molecular dynamics (MD). The potential energy surfaces expressed in the Morse form for Fe – N interactions are deduced from molecular DFT calculations and they allow producing Infra Red and Raman frequencies. These Fe – N potentials inserted in a classical force field lead from MD calculations to the relative energies of the high spin and low spin configurations of the orthorhombic structure. The MD investigations have also allowed assessing the experimental (P, T) phase diagram by showing the monoclinic polymorph in its two spin-states, and generating two triple points.*

**Keywords:** Spin-crossover complex; DFT; molecular dynamics; (P,T) phase diagram.

# 1. Introduction

Few inorganic transition metal ion complexes exhibit two electronic states of their *d* electrons, the Low Spin state (*LS*) and High Spin state (*HS*). The switching between these two states is subjected to small energy magnitudes and the transitions can be achieved with external constraints such as temperature and pressure as well as by applying light. The observed behavior, referred to as Spin CrossOver (*SCO*) and mainly studied in $Fe^{II}$ complexes, has been widely investigated by research groups worldwide both from the fundamental and applied (display applications, ...) [1-4]. The transition temperature at which there is same proportion of *LS* and *HS* is called $T_{1/2}$. In the case of first order transition, the thermal hysteresis is described by $T_{1/2}\uparrow$ and $T_{1/2}\downarrow$ associated with the corresponding enthalpy change $\Delta H$. One of theses complexes, *[Fe(PM-BIA)$_2$(NCS)$_2$]*[1], is characterized by iron and surrounded by a distorted octahedron involving two *N*-(2'-pyridylmethylene) (*PM*) and two 4-aminophenyl (*BIA*) ligands and completed by two thiocyanate anions. It is known to undergo a gradual *SCO* at $T_{1/2}$ = 190 K in the monoclinic polymorph (*II*-phase, *Z* = 4, space group *P2$_1$/c*) [1-8]. A very abrupt transition is observed for the orthorhombic polymorph (*I*-phase, *Z* = 4, space group *Pccn*) with $T_{1/2}\uparrow$ = 173 K and $T_{1/2}\downarrow$ = 168 K between the low spin (*LS*, S = 0) to the high spin state (*HS*, S = 2) of the $Fe^{2+}$ ion [4]. Among its interesting magneto-optical properties, this *SCO* may be light-induced, the limit temperature above which a photomagnetic effect in a material is erased - the so-called $T_{LIESST}$ (for « *Light-Induced Excited Spin-State Trapping* ») depending strongly of on the structural distortion of the $3d^6$ $Fe^{2+}$ environment [9] and/or the nature of the ligand [10]. As often [8,11,12], pressure plays a role in the opposite sense of temperature in as far the starting spin state is the high spin: by increasing pressure, both *I* and *II* polymorphs of *FePMBIA* yield a $HS \rightarrow LS$ transition [13].

---
[1] *[Fe(PM-BIA)$_2$(NCS)$_2$]* complex will be called thereafter as *FePMBIA*.

In such a complex compound, density functional theory (*DFT*) [14,15] approach is useful for understanding the property differences between the two spin states, namely ionic charges of the $Fe^{2+}$ cation or the *N* atoms, geometric molecular building and magnetic properties, as shown in the study of $Fe^{II}$ complexes: the *SCO* with a measurable $T_{1/2}$ originates from the crystal field around $Fe^{II}$ which must be of medium strength such with use of *N*-based ligands [16]. But *ab initio* methods cannot well explain intermolecular interactions in molecular crystals, because dispersion forces are insufficiently reproduced. Consequently, the obtained energy values may be far from experiments, leading sometimes to wrong hierarchy stability between the spin states. In view of this drawback the determination of the transition enthalpy $\Delta H_{LS \rightarrow HS}$ is made difficult.

For that, classical methods, such as molecular dynamics (*MD*), where van der Waals forces or hydrogen bonding may be modeled inside a generalized atom-atom force-field, are preferred. Nevertheless, the knowledge of atomic charges ($q_i$) and of the potential energy surface (*PES*) deduced from molecular *DFT* calculations, is very often used as a starting step in a *MD* approach, as in [17]; the quantum results allow evaluating force-field parameters. In this context we may define such an original methodology, combining *DFT* calculations and molecular dynamics, first in the molecular state and finally in the crystalline one, as "semi-classical molecular dynamics".

## 2. Methodology

Because the classical intramolecular force field comprises electrostatic forces acting between atomic charges, stretching 2-body interactions, bending 3-body forces and dihedral 4-body interactions, in order to model a molecule, the first step is a *DFT* procedure. *Ab initio* calculations applying the *GAUSSIAN* code to the *FePMBIA* molecule in its two spin states are firstly carried out, followed by a Mulliken analysis [18]. For this purpose, the hybrid *B3LYP\** functional, with effective core potential *LANL2DZ* (*Los Alamos National Laboratory with Double Zeta* polarization) function, has been used for all atoms. For a reminder, in the original formulation of the *B3LYP* hybrid functional, one modification of the exchange weighting parameter called *B3LYP\** was proposed by Reiher [19]. Resulting optimized molecular structure and point charges allow generating the *PES*. The curve is fitted with Morse $V_{Fe-N}$ and harmonic $V_{Fe-C-N}$ potentials, by varying alternatively one of the three *Fe-*$N_i$ distances of *FePMBIA* in each spin state while the two others remain constant. Only *I*-phase whose vibration properties are experimentally known has been used.

In a second step, *MD* simulations are performed including other intramolecular interactions described by 2-body, 3-body and 4-body potentials at the level of the molecule. Internal van der Waals interactions occurring between pyridine and phenyl rings or describing *S...H* hydrogen bonds must be added. All these parameters are adjusted with respect to the molecular geometry and the spectroscopic properties by using the *DL_POLY* code [20] which yields a molecular field for each spin state.

Then these generalized force fields can be applied to the molecular crystal lattice. After a minimization step at 1 K for both spin states, *MD* simulations are performed up to 300K by increasing temperature. Similar *MD* runs are done by cooling. Best van der Waals parameters are searched for, in order to reproduce structures both in *LS* state (25 K and 140K) and *HS* state (298 K) [2, 4, 21].

With these final intermolecular force fields, the complete set of runs allow evaluating respectively $T_{1/2}\uparrow$ and $T_{1/2}\downarrow$ and transition enthalpy $\Delta H_{LS \rightarrow HS}$ at $T_{1/2}$, the so-called $\Delta H_{1/2}$. $T_{1/2}$ and $\Delta H_{1/2}$ are related, as shown in a phenomological thermodynamical view, by modeling the *LS-HS* domain mixture as a regular solid-solution [22]. In such a way, the Gibbs free energy $G_{LS \rightarrow HS}$, referred to the *LS* state, is expressed as a function of the molar fraction of *HS* state $\gamma_{HS}$ in the crystal according to:

$$G_{LS \rightarrow HS} = \gamma_{HS} \Delta H_{LS \rightarrow HS} + \Omega \gamma_{HS}(1-\gamma_{HS}) + RT\{[\gamma_{HS} \ln \gamma_{HS} + (1-\gamma_{HS})\ln(1-\gamma_{HS})] - \Delta S_{LS \rightarrow HS}\}. \quad (1a)$$

$\Omega$ is a constant, which may be related to the interacting energy between domains ($\varepsilon_{LS-LS}$, $\varepsilon_{LS-HS}$, $\varepsilon_{HS-HS}$) and the domain size [23]. $\Delta S_{LS \rightarrow HS}$ represents the transition entropy during the spin transition, and *R* the ideal gas constant. Searching the equilibrium condition in the crystal as:

$$\partial G_{LS \rightarrow HS} / \partial \gamma_{HS} = 0, \quad (1b)$$

a relation ship is found in this model giving temperature as a function of thermodynamical functions:

$$T = [\Delta H_{LS \rightarrow HS} + \Omega(1-2\gamma_{HS})] / \{R \ln[(1-\gamma_{HS})/\gamma_{HS}] + \Delta S_{LS \rightarrow HS}\}, \quad (1c)$$

At the so-called $T_{1/2}$ temperature for which $\gamma_{HS} = 1/2$, eqn. (1c) implies that $T_{1/2}$ is $\Omega$-independent, as for an ideal mixture, and $T_{1/2} = \Delta H_{LS \rightarrow HS} / \Delta S_{LS \rightarrow HS}$. For other temperatures, $\gamma_{HS} \neq \frac{1}{2}$, even if $\Omega = 0$; the enthalpy function will be evaluated by using eqn. (1b) and assuming a Boltzman distribution of the *HS* domains. The variation enthalpy is then deduced as the enthalpic term of the Gibbs free energy:

$$\Delta H = \gamma_{HS} \Delta H_{LS \rightarrow HS} + \Omega \gamma_{HS}(1-\gamma_{HS}), \quad (1d)$$

with a *S*-shape, more pronounced when $\Omega \geq 2RT_{1/2}$.

At the same time, cell parameter variations versus temperature may be directly compared to experiments [3]. Isotherm and isobar *MD* runs are also performed on the *LS* state (*I*-phase) in order to study its instability, the resulting $I \rightarrow II$ transition in both spin states and to assess the experimental (*P,T*) phase diagram [13, 24].

## 3. Force field in the two spin states of a *FePMBIA* molecule

### 3.1. Potential energy surface fit

Fig.1 gives a molecular view of the *FePMBIA* complex. The numbering of the nitrogen atoms is the following: $N_1$ is attached to the pyridylmethylene group (*PM*), $N_2$ to the aminophenyl ring (*BIA*) and $N_3$ belongs to thiocyanate anion. *DFT* calculations are achieved for each spin state by a geometry optimization of this structure. The *PES* fit requires studying successively its sections by two of the *Fe-$N_1$*, *Fe-$N_2$* and *Fe-$N_3$* distances, in order to know its variation with *Fe-$N_1$*, *Fe-$N_2$* and *Fe-$N_3$* distances. Except the last case, the molecular geometry constraints imply that, when $d_{Fe-N1}$ distances vary, $\theta_{Fe-N2-C}$ angles are not constant. In the same manner, $\theta_{Fe-N1-C}$ angles vary with $d_{Fe-N2}$. Point Mulliken charges deduced from *DFT* calculations are given in Table I. For $Fe^{2+}$, they show that $q_{LS} \approx 3/2 \, q_{HS}$. Thus, it will not be surprising that existence of point Mulliken charges devoted to one spin state will imply two different intramolecular force fields.

All *Fe-$N_i$* (*i* = 1, … 3) 2-body interactions are described by a Morse expression:

$$V_{Morse}(r_{ij}) = E_{ij}^o \left[ \left\{ 1 - \exp\left(-\rho_{ij}\left(r_{ij} - r_{ij}^o\right)\right) \right\}^2 - 1 \right] \qquad (2)$$

where $E°_{ij}$ is the potential depth, $\rho_{ij}$ the electronic hardness and $r°_{ij}$ the equilibrium distance between two *i* and *j* species. It needs to be mentioned here that for the commodity of the many parameters for the six fits, we chose to make constant the $\rho_{ij}$ value to 1.33 Å$^{-1}$ independently from the nature of $N_i$ and of the spin state. A constant value of $\rho_{ij}$ is currently used in classical force field, even if the coordination polyhedra around a specific atom $i$ are of different types and strongly distorted (see as an example, [25]).

Furthermore we used harmonic 3-body interactions for the $Fe-N_1-C$ and $Fe-N_2-C$ angles:

$$V_{harm}(\theta_{ijk}) = \tfrac{1}{2} k_{ij} (\theta_{ijk} - \theta°_{ijk})^2 \qquad (3)$$

where $k_{ijk}$ is the stiffness constant of the angle $\theta_{ijk}$ and $\theta°_{ijk}$ the equilibrium angle.

Taking into account the Coulomb forces due to $q_{Fe2+}$ and $q_{Ni}$ charges, Figs. 2a-2c show the Morse fit of *PES* combined with angular harmonic potential (case of $N_1$ and $N_2$) when $Fe^{2+}$ is in the *LS* state. Figs. 3a-3c give the same kind of results for the *HS* state. All the corresponding parameters are given in Table II. It can be noted that the minimum position $r_{min}$ is smaller than $r°_{ij}$ as a consequence of the influence of the Coulomb field leading to a slightly modified curvature of the total potential $V(r_{ij}) = V_{Morse}(r_{ij}) + V_{Coulomb}(r_{ij})$ as seen by the 2$^{nd}$ order derivative:

$$\left[d^2V(r_{ij})/dr_{ij}^2\right]_{min} = 2E_{ij}^o \rho_{ij}^2 \exp\left[-2\rho_{ij}(r_{min} - r_{ij}^o)\right] + 2kq_iq_j(1 - 0.5 \times \rho_{ij}r_{min})/r_{min}^3 \qquad (4)$$

where $k$ is the Coulomb constant. From Table II, the potential depths in the *LS* state ($E^{o,LS}_{FeN1}$, $E^{o,LS}_{FeN2}$, $E^{o,LS}_{FeN3}$) are deeper than in the *HS* state ($E^{o,HS}_{FeN1}$, $E^{o,HS}_{FeN2}$, $E^{o,HS}_{FeN3}$) and :

$$E^{o,LS}_{FeN1} \sim 3E^{o,HS}_{FeN1} \text{ and } E^{o,LS}_{FeN2} \sim 3E^{o,HS}_{FeN2}.$$

### 3.2. Intramolecular force field

The force field devoted to a given spin state requires the input of:

i. all 2-body stretching harmonic interactions, such as *C-H*, *C-C*, *C-N*, *C-S*, *N-S*, but also weak *N-N* octahedron interactions;

*ii.* 3-body interactions, namely *H-C-C*, *H-C-N*, *C-C-C*, *C-C-N*, which will be described by bending harmonic potentials associated (eqn. (2)) and coupled potentials. There are two extra contributions needing to be accounted for in the coupling potentials; these are the coupling between adjacent bonds for species *i,j* and *i,k* in $V(r_{ij}, r_{ik})$:

$$V(r_{ij}, r_{kj}) = A_{ijk} (r_{ij} - r°_{ij})(r_{ik} - r°_{ik}) \qquad (5)$$

and the potential accounting for the bond and the angle $V(r_{ij}, \theta_{ijk})$:

$$V(r_{ij}, \theta_{jk}) = A_{ijk} (r_{ij} - r°_{ij})(\theta_{ijk} - \theta°_{ijk}); \qquad (6)$$

in booth eqn. (4)-(5) $A_{ijk}$ is a stiffness constant.

*iii.* 4-body dihedral and inversion interactions for maintaining the planarity of pyridine and of phenyl rings as well as for describing low vibration modes. The 4-body dihedral potential is:

$$V_{cos} = A_{ijkl} [1 + \cos(m\phi_{ijkl} - \delta_{ijkl})] \qquad (7)$$

where $A_{ijkl}$ is a stiffness constant, $\phi_{ijkl}$ is the dihedral angle and $\delta_{ijkl}$ the reference dihedral angle acting as an out-of-phase. Interactions due to inversion (with the $\phi_{ijkl}$ angle) correspond to the following expression:

$$V_{plan} = A_{ijkl} [1 - \cos(\phi_{ijkl})]. \qquad (8)$$

Starting values for parameters involved in all these intramolecular potentials, i.e. equations (2) - (3), (5) - (8), come from the *COMPASS* force field which is devoted to organic molecules [26].

*iv.* long-range potentials in a Lennard-Jones form $V_{LJ}(r_{ij})$ have to be accounted for to discribe the van der Waals interactions between adjacent phenyl rings and *S...H* hydrogen bonds and to reproduce the deformation modes involving two ligands connected to $Fe^{2+}$:

$$V_{LJ}(r_{ij}) = 4\,\varepsilon^o_{ij}\,[(\sigma_{ij}/r_{ij})^{12} - (\sigma_{ij}/r_{ij})^6]; \tag{9}$$

$\varepsilon^o_{ij}$ is the potential depth, $\sigma_{ij}$ the distance at which the long-rang $i$-$j$ interaction vanishes. The corresponding parameter values are extracted from [27].

The $r^o_{ij}$ equilibrium Morse values are changed in order to match the *MD* geometry ($r_{min}$) with experiment. In the same way the stiffness constants in the other 2-body potentials and in the 3-body potentials, as well as in the coupled ones, are modified according to the vibration modes. For both spin states, the complete force fields are given as supplementary informations (Tables S1 and S2).

### 3.3. *Application to the molecule in LS and HS spins states*

The established force fields are applied to the *FePMBIA*-molecule in *MD* runs by using *DL_POLY* code. All simulations are performed at 0 K in the *NVT* ensemble coupled to a Berendsen thermostat (relaxation time constant = 0.1 ps) [28], with a long-range distance cut-off equal to 12 Å when calculating Coulomb forces by the Ewald method [29] or van der Waals interactions. As a result, Table III provides a comparison between experimental and *MD* octahedral $FeN_6$ geometries for the two spin states. The $d_{Fe-N}$ distance magnitudes are well reproduced together with the $\theta_{N-Fe-N}$ angles. The expected trend for a greater distortion in *HS* octahedron versus *LS* is found as exhibited by distance and angle extrema:

$$\Delta d_{HS} = 0.21 > \Delta d_{LS} = 0.03 \text{ Å}; \; \Delta\theta_{HS} = 26.4 > \Delta\theta_{LS} = 13.7 \text{ deg.}$$

Low-frequency vibration modes are well reproduced. Table IV shows that the difference $\Delta\nu_{MD}$ between measurements of Hoeffer [30] and *MD* results does not exceed 23 cm$^{-1}$ for octahedral

*Fe-N* stretching modes, whereas *DFT* agreement is better ($\Delta v_{DFT} < 12$ cm$^{-1}$). The position of the bending modes of the *N-C-S* group appear below experimental values ($\Delta v_{MD} < 35$ cm$^{-1}$ in *LS*-state, $< 10$ cm$^{-1}$ in *HS*-state); agreement is better for stretching *C-S* and *C-N* modes in both states ($\Delta v_{MD} < 24$ cm$^{-1}$). In the high-frequency region due to the *C-H* modes, well matching values are found upon comparing with *DFT* calculations; for these modes, $0 < v_{LS} - v_{HS} < 49$ cm$^{-1}$, smaller than with *DFT* values (Table S3 as supplementary informations).

For each spin state, the total energy found at 0 K ($E_{LS}$ = -26.11 eV, $E_{HS}$ = -25.99 eV) yields directly the molecular total energy difference between the two spin states $\Delta E_{LS \rightarrow HS}$:.

$$\Delta H_{LS \rightarrow HS} = \Delta E_{LS \rightarrow HS} + P\Delta V_{LS \rightarrow HS}. \qquad (10)$$

Because MD runs are done under the *NVT* ensemble, $\Delta V_{LS \rightarrow HS} = 0$ in this expression, and therefore, $\Delta E_{LS \rightarrow HS}$ gives directly the transition enthalpy: $\Delta H_{LS \rightarrow HS} = 0.12$ eV $= 11.2$ kJ/mol. Incidently, this value is close to the experimental one determined by differential scanning analysis (*DSC*) on *FePMBIA* crystalline powders, but obtained in the 160-190 K range, by [2] ($\Delta H_{LS \rightarrow HS} = 10.06$ kJ/mol) and more recently by [31] ($\Delta H_{LS \rightarrow HS} = 11.5$ kJ/mol).

**4. *I-FePMBIA* crystal in the two spin states**

4.1. *Intermolecular field in the crystal phase*

At this step, molecular force field in one of the two spin states may be applied in *MD* runs to the crystal. For this purpose, the experimental structure corresponding with its three cell parameters (*a*, *b* and *c*), determined at 25 K and at room-temperature, is used as a starting configuration and repeated in each crystallographic direction. The resulting 8-cell « box » is optimized at 0 K in the *NVT* ensemble coupled to a Berendsen thermostat (relaxation time constant = 0.1 ps) with the same cut-off as in section 3.3 (12 Å). Then, isotherm simulations are made in the isostress $N\sigma T$ ensemble coupled to a Berendsen barostat (relaxation time constant = 1.0 ps): *i)* to ensure that the average system pressure is maintained to the atmospheric one; *ii)* to allow $b/a$ and $c/a$ ratios to vary; *iii)* to ensure that the orthorhombic symmetry is conserved (the angles α, β and γ of the simulated cell ought to be very close to 90°). Contrary to an isobar simulation run performed in the *NPT* ensemble, isostress *NσT* ensemble implies that pressure does not rigorously conserve the angles within the cell. Each run is followed by a new simulation in the isobar *NPT* ensemble with the same relaxation time constants as in the *NVT* or *NσT* runs.

It is clear that cell volume and structure will be sensitive to the most numerous van der Waals interactions, that is $C…C$, $C…H$ and $H…H$ interactions, but also to the more energetic hydrogen bonds ($N…H$ and $S…H$). Without compressibility data, it is impossible to adjust all the corresponding Lenard-Jones potentials (eqn. (9)). Therefore, we introduce the constraint that each potential depth $\varepsilon°_{ij}$ does not depend on the spin state and we modify $C...C$, $C...N$, $C...S$, $N...N$, $N...S$, $N...H$ and $S...H$ parameters by looking for their influence on the crystal data at 25 K (*LS*-state) and 298 K (*HS*-state) and on the corresponding van der Waals contacts. Table V gives the complete set of parameters ($\varepsilon°_{ij}$ and $\sigma_{ij}$) for both spin states. Between the *LS*- and *HS*- states, some differences may be noted, mainly for $C...S$ and $S...H$ interactions where respectively $^{LS}\sigma_{C…S}$ = 3.040 Å is smaller than $^{HS}\sigma_{C…S}$ = 3.741 Å and $^{LS}\sigma_{S…H}$ = 2.470 Å smaller than $^{HS}\sigma_{S…H}$ = 2.900 Å: this means that $S...H$ hydrogen bonds are stronger in the *LS*-phase than in the *HS*-phase, in agreement with the evolution of corresponding intermolecular distances with $T$ ($^{LS}d_{C…S}$ = 3.796 Å, $^{HS}d_{C…S}$ = 3.901 Å, and $^{LS}d_{S…H}$ = 2.94 Å, $^{HS}d_{S…H}$ = 3.03 Å), given by [4, 32]. The van der Waals part of the energy $E_{vdw}$ represents a little more of the 1/10 of the total energy ($^{LS}E_{vdw}$ = -281.25 kJ/mol, $^{HS}E_{vdw}$ = - 278.90 kJ/mol); but if these intermolecular terms are disregarded, $\Delta H_{LS \to HS}$ will be underestimated of ~ 2.3 kJ/mol.

### 4.2. *Intermolecular interactions in the crystal state: towards thermoinduced transition LS→HS*

After this van der Waals interaction determination, we are able to simulate an orthorhombic phase cell in its two spin states at 0 K under atmospheric pressure: the total *HS* energy ($E_{HS}$ = - 28.65 eV = -2674.0 kJ/mol) is found to be slightly lower than in the *LS* state ($E_{LS}$ = -28.29 eV = -2640.4 kJ/mol). By constraining the van der Waals potential depths to be independent of the spin state, these total energy values are certainly not correct. Morever, quantum effects prevail at low temperatures. At theses temperatures, because the difference between these values is of the order of their numerical fluctuation (± 10 kJ/mol), the transition enthalpy cannot be evaluated precisely through an equation similar to eqn. (10), even if *DL_POLY* code gives us both total energy *E* and enthalpy *H* during each step of a *NPT* or *NσT* run.

A better value for $\Delta^{MD} H_{LS \to HS}$ must be determined in another manner. For this purpose, the molecular crystal system has to be studied by varying its thermodynamic state through isotherm-isobaric runs. Increasing temperature runs are made on *LS*-phase at 0 K minimized by applying *LS*-field; this field is also used for the resulting room temperature phase but with decreasing temperature, in order to study the hysteresis behavior near the $T_{1/2}$ temperature. Similar approach is used for *HS*-phase starting in the 0-298 K temperature range with increasing and decreasing step runs.

Figs. 4a-4c give the respective variation of *a*, *b* and *c* cell parameters as a function of *T*. The three curve sets show different behaviors. The *a*-curve, obtained when *HS*-force field is applied to the *HS*-phase, shows the instability of this phase below 100K, whereas the *c*-curves

is too imprecise for extracting transition informations. However, the crossing of the *b*-curves, obtained with *LS*- and *HS*-force fields and applied with increasing temperature, specifies more precisely the $T_{1/2}\uparrow$ temperature ($T_{1/2}\uparrow$ = 120 K), which is 50 K lower than the experimental one (173 K, [2]). As a result, the effect of the instability of the *HS*-state below 100K, shown in the evolution of the *a* cell parameter, is found again in the cell-volume: by increasing temperature a cell-volume expansion occurs at 100 K (Fig.4d). At this temperature, the difference of the cell volume between *LS* and *HS*-phases ($\Delta V_{cell}$ = 121 Å$^3$) may be compared to the experimental value deduced form crystal data ($\Delta V_{cell}$ = 57.4 Å$^3$, [3]). The observed relatively large differences in temperature and in volume can be accounted for by the drawbacks of a purely classical model in view of the prevailing quantum effects in this temperature region.

At each run step, the total energy and therefore the enthalpy given by the *DL_POLY* code are spin state dependent. The enthalpy difference $\Delta^{MD}H_{LS\rightarrow HS}$ between the two states values is obtained by acting the *LS*-field on *LS* and *HS* configurations at increasing temperature and reported on Fig.5 (lower points). A sharp transition occurs between 140 and 160K, when $\gamma_{HS}$ content increases according to eqn.(1a-1c). By using now the *HS*-field on both spin state phases during cooling runs (Fig.5, upper points), $\Delta^{MD}H_{LS\rightarrow HS}$ is found not to depart from 14±1 kJ/mol, which corresponds to the absence of *LS* crystal domains ($\gamma_{HS}$=1). Therefore, $\Delta^{MD}H_{1/2}$ may be taken as the value of $\Delta^{MD}H_{LS\rightarrow HS}$, that is $\Delta^{MD}H_{1/2}$ ~14±2 kJ/mol (the slightly lower accuracy is due to the fluctuations of the *LS* background around zero). This *MD*-value lies a little above the data arising from *DSC* ($\Delta^{DSC}H_{LS\rightarrow HS}$ = 10.06 kJ/mol, [2]; 11.5 kJ/mol, [31]). A more precise determination of $\Delta^{MD}H_{1/2}$ would be certainly obtained through the evaluation of the Gibbs free energy variation $\Delta^{MD}G_{LS\rightarrow HS}$ resulting from the integration of the hybrid classical Hamiltonian which contains potential and kinetic terms and which describes both *LS* and *HS* spin states (see [20, 27]). Nevertheless, the evolution of $\Delta^{MD}H_{LS \to HS}$ with *T* leads to a transition temperature value ($T_{1/2}$ =150 K), more precise and higher than the one deduced from the structural results (see above), and therefore in better agreement when compared to the experiments ($T_{1/2}\uparrow$= 173 K and $T_{1/2}\downarrow$ = 168 K, [2]).

The results of simulations presented here allow also comparisons with experimental structures (25 K, 140 K and 298 K). Then, in the *LS* spin state, Table VI shows that the *b* parameter is larger by 7.2 % at 25 K and 12.5 % at 140 K, but this is compensated by a *c* parameter with a lower value (-11.7 % at 140 K). This is even better shown for the *HS* spin state where the small differences on the three cell parameters compensate perfectly when looking for the $V_{cell}$ cell volume: $\Delta V_{cell}/V_{cell}$ = 0.7 %. All this result set yields to prefer the more precise thermodynamical value $T_{1/2}$ = 150 K.

## 5. Pressure stability of *I-FePMBIA* crystal: the pressure induced *I* → *II* phase transition

The influence of pressure on *FePMBIA* has been studied by magnetic susceptibility measurements [13], optical reflectivity [8] and completed by a recent neutron diffraction study [24]: the orthorhombic *I*-phase undergoes a transition towards the monoclinic *II*-phase, independtly of the spin state. This expectedly means that the (*P,T*) phase diagram must involve four solid phases $LS_I$, $HS_I$, $LS_{II}$ and $HS_{II}$, and therefore two triple points between three of these four phases. Because it would be difficult to study experimentally these last parts of the (*P,T*) phase diagram, in particular the respective positions of the triple points, *MD* is helpful for assessing it. Therefore, simulations have been performed in the *NσT* ensemble in different isotherm-isobar conditions from 25 to 300 K with pressure up to 20 kbar. All these processes were carried out for orthorhombic *[Fe(PM-BIA)$_2$(NCS)$_2$]* in order to exhibit its instability with increasing pressure. Furthermore, simulating the monoclinic phase in both spin states would have required establishing not only a new intramolecular parameter set, but also to select appropriate van der Waals force field in *II*. It is well admitted to use the same force field for room pressure and high pressure phases, as in $B_2O_3$ [25]. Because of the symmetry breaking between *I* and *II* structures, the corresponding phase transition for each spin state will be of the first order and therefore better detected in modeling procedures. At low temperature, the cell volumes of *I* and *II* in their *LS* state are known to be very different ($V_{cell, LSI}$ = 3339 Å$^3$ at 140 K, $V_{cell, LSII}$ = 3294 Å$^3$ in metastable state at 120 K, [4]); that means increasing pressure will favor more strongly the $LS_I \rightarrow LS_{II}$ phase transition.

Figs. 6a-6c give the respective variations of *a*, *b* and *c* cell parameters as a function of *P* for different isotherms between 25 K and 160 K, when $LS_I$ force field is acting: with increasing *P*, *a* and *c* show a suddenly decreasing in the range 7-12 kbar for the 25 K, 50 K, 80 K, whereas a small *b*-jump is observed; a departure of the *β* angle from 90° increases. Because the resulting volume variation is negative, *SCO* must not occur and therefore, as argued before, a $LS_I \rightarrow LS_{II}$ transition is assumed. For upper isotherms, the *a* and *c* curves have smaller discontinuities, one at lower *P*-values, namely around 3 kbar, and the second one at 9 kbar. By comparison with the experiments of [13], two transitions must be taken into account: *(i)* (i) for the lower transition, by combining *SCO* with a change of symmetry, pressure may authorize a phase in the *HS* state, but more compact than the $LS_I$ polymorph (here $\Delta V_{cell}$ = -35 Å$^3$), leading to a $LS_I \rightarrow HS_{II}$ transition ; *(ii)* because pressure will favor a transition from *HS* to *LS* state, the upper transition will be $HS_{II} \rightarrow LS_{II}$.

When $HS_I$ force field is applied during *MD* runs, $HS_I$ phase will transform logically into a monoclinic $HS_{II}$ polymorph. Figs. 7a-7c give the respective variations of *a*, *b* and *c* cell parameters as a function of *P* for different isotherms between 190 K and 300 K : the discontinuities appearing on the *a* and *b* curves, leading to a cell volume contraction, show that the expected $HS_I \rightarrow HS_{II}$ transition occurs at $P \approx 3$ kbar below 280 K, at significantly higher pressure at 300 K ($P \approx 5$ kbar where $\Delta V_{cell} = -70$ Å$^3$).

Knowing both enthalpy and volume variations during the $LS_I \rightarrow HS_I$ transition at $T_{1/2}$, as given in section 4.2 ($\Delta H_{1/2} = 14.0$ kJ/mol, $\Delta V_{cell} = 121$ Å$^3$), the use of the Clausius-Clapeyron relation in its form valid for the transitions of the first kind:

$$(dP/dT)_{T1/2} = (\Delta H_{LSI \rightarrow HSI}) / (T\Delta V_{LSI \rightarrow HSI}) \qquad (11)$$

allows estimating the slope of the curve corresponding to the $LS_I \rightarrow HS_I$ transition at 1 bar: $(dP/dT)_{T1/2} = 0.08$ kbar/K. Therefore, eqn.(11) may be used as a guide for the curve delimiting $LS_I$ and $HS_I$ area. One may remark that the value found is 1/2 of the one deduced from the experimental slope of a straight line separating $LS_I$ and $HS_I$ regions built from data of [24] (($dP/dT)_{T1/2} = 0.16$ kbar/K). Nevertheless, this difference is acceptable if we keep in mind the uncertainties found in the evaluation of $\Delta H_{1/2}$, $T_{1/2}$ and $\Delta V_{cell}$. Moreover, the present *MD*-value of $(dP/dT)_{T1/2}$ is of the same order to the one found for the [Fe(sal2-trien)][Ni(dmit)2] complex which is known to have its *SCO* near 245 K (($dP/dT)_{T1/2} = 0.06$ kbar/K, [11]).

From an experimental point of view, by putting the respective values of $T_{1/2}$ (170K) and $\Delta V_{cell}$ (57.4 Å$^3$) and knowing the slope $dP/dT$ (0.16 kbar/K), $\Delta H_{1/2}$ can be evaluated from eqn.(11) ($\Delta^{CC}H_{1/2} = 23.5$ kJ/mol). This value is twice the *DSC* one ($\Delta^{DSC}H_{LS \rightarrow HS} = 10.06$ kJ/mol [2], 11.5 kJ/mol [31]). In fact, this procedure is certainly incorrect, because in the pressure range involved (1 bar to 6 kbar) the true limiting curve between $LS_I$ and $HS_I$ domains is probably not a straight line, as a consequence of the evolution of the compressibility with pressure in each phase. This means that the slope $dP/dT$ should be smaller than 0.16 kJ/mol, and therefore, the deduced $\Delta^{CC}H_{1/2}$ value closer to the DSC ones.

At this step, the (*P,T*) phase diagram may be built and compared to experiments (Fig.9). *MD* points are shifted to lower pressure and to lower temperature. Consequently, the calculated triple points are more separated than the ones estimated from experiments: the low pressure triple point which defines the $LS_I$, $HS_I$ and $HS_{II}$ equilibrium is located at 170 K and a pressure of ~ 3.0 kbar, the second triple point corresponding to the $LS_I$, $LS_{II}$ and $HS_{II}$ equilibrium is close to 100 K at ~ 6.8 kbar. These results may be compared to the experiments [13]. By studying the pressure effect on the hysteresis loop, these authors concluded that the data in the 6-8 kbar range correspond to a *SCO* coupled to a crystallographic phase transition. The present investigation confirms these findings and further provides the nature of the spins states according to the natures of the polymorphs, *I* and *II*. In other words, because the data of ref. [13] take into account the metastability of the $LS_I$ phase in the corresponding thermodynamic conditions which transforms into the $HS_{II}$ phase, the line delimiting $HS_I$ and $HS_{II}$ phase domains in Fig.9 would be located at pressures lower than 7 kbar and at lower temperatures, as found by *MD*.

Another interest in this phase diagram is found by inspecting the slopes of the curves corresponding to the two transitions $LS_I \rightarrow LS_{II}$ and $HS_I \rightarrow HS_{II}$ which are on opposite directions: $(dP/dT)_{LSI \rightarrow LSII}$ is clearly negative, whereas $(dP/dT)_{HSI \rightarrow HSII}$ is slightly positive below room temperature. Because pressure favors more compact *II*-phases ($LS_{II}$ or $HS_{II}$) with a smaller molar volume as shown by [4] ($V_{cell, LSII}$ = 3294 Å$^3$ < $V_{cell, LSI}$ = 3399 Å$^3$), eqn. (11) implies that pressure and temperature play antagonist roles during the $LS_I \rightarrow LS_{II}$ transition with a positive enthalpy variation $\Delta H_{LSI \rightarrow LSII}$. On the other hand, the $HS_I \rightarrow HS_{II}$ transition will correspond to a small negative enthalpy variation $\Delta H_{HSI \rightarrow HSII}$, associated, or not, to a small cell volume variation at room-temperature, and to a significant cell volume contraction for *T* below 280 K. This last assumption is confirmed by the behavior of the corresponding experimental cell volumes [4, 24, 32]: starting from nearly equal values at 293 K ($V_{cell, HSII}$ = 3464 Å$^3$; $V_{cell, HSI}$ = 3462 Å$^3$), the two volume values diverge with decreasing temperature ($\Delta V_{cell}$ = -20 Å$^3$ at 250 K, $\Delta V_{cell}$ = -30 Å$^3$ at 225 K), until 210 K where $HS_{II}$-phase becomes unstable ($\Delta V_{cell}$ = -40 Å$^3$).

It will kept in mind that all discrepancies appearing in this section arise partly from assuming one force field for *I* and *II*, in particular for long-range van der Waals interactions. Nevertheless, the observed trends should not be changed.

## 5. Conclusions

Starting from the potential energy surface of the *[Fe(PM-BIA)$_2$(NCS)$_2$]* complex obtained by DFT calculations on its distorted molecular geometry, Morse two-body *Fe-N* interactions have been evaluated for both spin states of the orthorhombic polymorph (*I*). The potential wells and the equilibrium distances depend both of the nature of the bound *N*-atom and of the spin state of the *Fe$^{2+}$* ion. An adjustment of other two-body potentials, combined with van der Waals interactions, allow then to calculate vibration modes (I.R. and Raman) of the molecule in its two spin states by molecular simulation. In agreement with the experimental molecular geometry, the *FeN$_6$* octahedron is found more distorted in the high-spin state than in the low-spin one, as exhibited by distance ($\Delta d_{HS}$ = 0.21 > $\Delta d_{LS}$ = 0.03 Å) and angle extrema ($\Delta \theta_{HS}$ = 26.4 > $\Delta \theta_{LS}$ = 13.7 deg). Transferring the two obtained fields to the crystal state, molecular simulations give structural informations on the crystal lattice of the orthorhombic phases. Intermolecular interactions show evidence of hydrogen bonding between *NCS* sulfur end atoms with nearest neighbors hydrogen belonging to the aromatic cycles. The evolution of the structure with temperature shows that the *LS → HS* transition occurs at $T_{1/2}\uparrow$ = 120 K which is 50K lower than the experimental value. The corresponding volume change of the unit cell, $\Delta V_{cell} = 121$ Å$^3$, may be compared to the crystallographic data at 170K ($\Delta V_{cell} = 57.4$ Å$^3$).

The change of the transition enthalpy, $\Delta^{MD}H_{1/2} = 14$ kJ/mol corresponds to the transition temperature $T_{1/2} = 150$ K. If $T_{1/2}$ agrees with experiments ($T_{1/2}\uparrow = 173$ K and $T_{1/2}\downarrow = 168$ K), the MD enthalpy value lies a little above experimental measurements ($\Delta^{DSC}H_{LS \to HS} = 10.06$ kJ/mol and 11.5 kJ/mol). A more precise determination of $\Delta^{MD}H_{1/2}$ would certainly be obtained through the evaluation of the Gibbs free energy variation $\Delta^{MD}G_{LS \to HS}$ resulting from the integration of the hybrid classical Hamiltonian which contains potential and kinetic terms and which describes both LS and HS spin states.

The relative disagreement for $\Delta^{MD}H_{1/2}$ and for the structural results can also be accounted for by the drawbacks of a purely classical model in view of the prevailing quantum effects accounting below 100 K.

The molecular simulations have led to assess and complete the experimental (P,T) phase diagram of *[Fe(PM-BIA)$_2$(NCS)$_2$]*. This is relevant to showing the domains of existence of the monoclinic polymorph (*II*) generated by its two spin states. Furthermore the delimitation of the different domains allowed generating two triple points:

- for $LS_I$, $HS_I$ and $HS_{II}$ equilibrium: at 170 K and ~ 3.0 kbar,
- for $LS_I$, $LS_{II}$ and $HS_{II}$ equilibrium, at 100 K and ~ 6.8 kbar.

On the other hand, the lack of data at low temperature leads us to provide neutron diffraction experiments on the $LS_I$ phase, in order to precise the (P,T) phase diagram below 175 K, in particular the stability range of this phase.

Indirectly, this work contributes to the general efforts done to identify the factors allowing to control the existence of the *SCO* phenomenon in molecular materials. This demonstrates how the packing may affect the geometry of a ligand and then the *SCO* properties. All the present results would be improved by using specific force field derived for each one of the four phases. This implies obtaining new van der Waals force field parameters as a function of the spin state and of the symmetry.

# REFERENCES

[1] J.F. Létard, S. Montant, P. Guionneau, P. Martin, A. Le Calvez, E. Freysz, D. Chasseau, R. Lapouyade, O. Kahn, *J. Chem. Soc., Chem. Comm.*, 1997, 745 ; J. Degert, N. Lascoux, S. Montant, S. Létard, E. Freysz, G. Chastenet, J.-F. Létard, *Chem. Phys. Lett. Lett.*, 2005, **415**, 206.

[2] J.F. Létard, P. Guionneau, L. Rabardel, J.A.K. Howard, A.E. Goeta, D. Chasseau, O. Kahn, *Inorg. Chem.*, 1998, **37**, 4432.

[3] H. Daubric, J. Kliava, P. Guionneau, D. Chasseau, J.F. Létard, O. Kahn, *J. Phys. Condens. Matter*, 2000, **12**, 5481.

[4] M. Marchivie, P. Guionneau, J.F. Létard, D. Chasseau, *Acta Cryst. B*, 2003, **59**, 479.

[5] J.-F. Létard, G. Chastanet, O. Nguyen, S. Marcen, M. Marchivie, P. Guionneau, D. Chasseau, P. Gütlich, *Monatshefte für Chemie*, 2003, **134**, 165.

[6] Ichiyanagi, J. Hébert, L. Toupet, H. Cailleau, E. Collet, P. Guionneau, J.-F. Létard, *Solid State Phenomena*, 2006, **112**, 81.

[7] M. Buron-Le Cointe, J. Hébert, C. Baldé, N. Moisan, L. Toupet, P. Guionneau, J. F. Létard, E. Freysz, H. Cailleau, E. Collet, *Phys. Rev. B*, 2012, **85**, 064114.

[8] A. Rotaru, F. Varret, E. Codjovi, K. Boukheddaden, J. Linares, A. Stancu, P. Guionneau, J.-F. Létard, *J. Appl. Phys.*, 2009, **106**, 053515.

[9] M. Marchivie, P. Guionneau, J.F. Létard, D. Chasseau, *Acta Cryst. B*, 2005, **61**, 25.

[10] J.-F. Létard, *J. Mater. Chem.*, 2006, **16**, 2550.

[11] P.A.Szilagyi, S. Dorbes, G. Molnar, J.A. Real, Z. Homonnay, C. Faulmann, A. Bousseksou, *J. Phys. Chem. Solids*, 2008, **69**, 2681.

[12] H.J. Shepherd, P. Rosa, L. Vendier, N. Casati, J.F. Létard, A. Bousseksou, P. Guionneau, G. Molnr, *Phys. Chem. Chem. Phys.*, 2012, **14**, 5265.

[13] V. Ksenofontov, G. Levchenko, H. Spiering, P. Gütlich, J.F. Létard, Y. Bouhedja, O. Kahn, *Chem. Phys. Lett.*, 1998, **294**, 545.

[14] P. Hohenberg, W. Kohn, *Phys. Rev. B*, 1964, **136**, 864.

[15] W. Kohn, L.J. Sham, *Phys. Rev. A*, 1965, **140**, 1133.

[16] L. Kabalan, S.F. Matar, *Chem. Phys.*, 2009, **359**, 14; S.F. Matar, J.F. Létard, *Z. Naturf. B*, 2010, **65b**, 565.

[17] Y. Zhu, Y. Su, X. Li, Y. Wang, G. Chen, *Chem. Phys. Lett.*, 2008, **455**, 354.

[18] *GAUSSIAN* 03, Revision C.02, M.J. Frisch, G.W. Trucks, H.B. Schlegel, G.E. Scuseria, M.A. Robb, J.R. Cheeseman, J.A. Montgomery Jr, T. Vreven, K.N. Kudin, J.C. Burant, J.M. Millam, S.S. Iyengar, J. Tomasi, V. Barone, B. Mennucci, M. Cossi, G. Scalmani, N. Rega, G.A. Petersson, H. Nakatsuji, M. Hada, M. Ehara, K. Toyota, R. Fukuda, J. Hasegawa, M. Ishida, T. Nakajima, Y. Honda, O. Kitao, H. Nakai, M. Klene, X. Li, J.E. Knox, H.P. Hratchian, J.B. Cross, V. Bakken, C. Adamo, J. Jaramillo, R. Gomperts, R.E. Stratmann, O. Yazyev, A.J. Austin, R. Cammi, C. Pomelli, J.W. Ochterski, P.Y. Ayala, K. Morokuma, G.A. Voth, P. Salvador, J.J.D. Dannenberg, V.G. Zakrzewski, S. Dapprich, A.D. Daniels, M.C. Strain, O. Farkas, D.K. Malick, A.D. Rabuck, K. Raghavachari, J.B. Foresman, J.V. Ortiz, Q. Cui, A.G. Baboul, S. Clifford, J. Cioslowski, B.B. Stefanov, G. Liu, A. Liashenko, P. Piskorz, I. Komaromi, R.L. Martin, D.J. Fox, T. Keith, M.A. Al-Laham, C.Y. Peng, A. Nanayakkara,

M. Challacombe, P.M.W. Gill, B. Johnson, W. Chen, M.W. Wong, C. Gonzalez, J.A. Pople, Gaussian, Inc., Wallingford CT, 2004.

[19] M. Reiher, *Inorg. Chem.*, 2002, **41**, 6928.

[20] *DL_POLY* 2.20, W. Smith, T.R. Forester, I.T. Todorov, SFTC Daresbury Laboratory, Warrington, UK, 2009.

[21] P. Guionneau, C. Brigouleix, Y. Barrans, A.E. Goeta, J.F. Létard, J.A.K. Howard, J. Gaultier, D. Chasseau, *C.R. Acad. Sci. Ser. IIc*, 2001, **4**, 161.

[22] C.P. Slichter, H.G. Drickamer, *J. Chem. Phys.*, 1972, **56**, 2142.

[23] C. Cantin, J. Kliava, A. Marbeuf, D. Mikaïlitchenko, *Eur. Phys. J. B*, 1999, **12**, 525.

[24] V. Legrand, F. Le Gac, P. Guionneau, J.F. Létard, *J. Appl. Cryst.*, 2008, **41**, 637.

[25] A. Takada, C.R.A. Catlow, G.D. Price, *J. Phys. Condens. Matter.*, 1995, **7**, 8659.

[26] H. Sun, *J. Phys. Chem. B*, 1998, **102**, 7338; S.W. Bunte, H. Sun, *J. Phys. Chem. B*, 2000, **104**, 2477; J. Yang, Y. Ren, A. Tian, H. Sun, *J. Phys. Chem. B*, 2000, **104**, 4951; M.J. McQuaid, H. Sun, D. Rigby, *J. of Comput. Chem.*, 2004, **25**, 61.

[27] P. Cazade, *Thèse*, Université de Pau et des Pays de l'Adour, 2008.

[28] H.J.C. Berendsen, J.P.M. Postma, W. van Gunsteren, A. DiNola, J.R.J. Haak, *J. Chem. Phys.*, 1984, **81**, 3684.

[29] M.P. Allen, D.J. Tildesley, *Computer Simulation of Liquids*, Clarendon Press, Oxford, 1989.

[30] A. Hoeffer, *Thesis*, Mainz Universität, 2000.

[31] D. Mondieig, P. Négrier, P. Guionneau, 2008, internal report, Université de Bordeaux.

[32] P. Guionneau, M. Marchivie, G. Bravic, J.F. Létard, D. Chasseau, *J. Mater. Chem.,* 2002, **12**, 2546.

Table I.- Mulliken charges of atoms in the $Fe$-$(N)_6$ octahedron (in e).

|        | LS      | HS      |
|--------|---------|---------|
| $Fe$   | 1.5099  | 1.0442  |
| $N_1$  | -0.7245 | -0.7662 |
| $N_2$  | -0.7678 | -0.6936 |
| $N_3$  | -0.7567 | -0.7326 |

Table II.- Parameters for Morse potentials describing the 2-body $Fe$-$N_i$ interactions and harmonic potentials of 3-body $Fe$-$N_i$-$C$ interactions deduced from the $DFT$ potential energy surface.

|         | $E^o$ (eV) | $\rho$ (Å$^{-1}$) | $r^o$ (Å) | $r_{min}^{(opt)}$ (Å) | $r_{min}^{(DFT)}$ (Å) | $k$ (meV/deg$^2$) | $\theta_{min}^{(opt)}$ (deg) | $\theta_{min}^{(DFT)}$ (deg) |
|---------|------------|-------------------|-----------|-----------------------|-----------------------|-------------------|------------------------------|------------------------------|
| LS      |            |                   |           |                       |                       |                   |                              |                              |
| $Fe$-$N_1$ | 1.54(1)  | 1.33              | 2.34(1)   | 2.000(5)              | 1.998                 | 1.0               | 119(1)                       | 114.5                        |
| $Fe$-$N_2$ | 1.33(1)  | 1.33              | 2.38(1)   | 1.990(5)              | 1.983                 | 0.52              | 119(1)                       | 114.3                        |
| $Fe$-$N_3$ | 1.23(2)  | 1.33              | 2.355(5)  | 1.925(5)              | 1.936                 |                   |                              |                              |
| HS      |            |                   |           |                       |                       |                   |                              |                              |
| $Fe$-$N_1$ | 0.44(2)  | 1.33              | 2.78(1)   | 2.280(5)              | 2.28                  | 0.52              | 119(1)                       | 117.0                        |
| $Fe$-$N_2$ | 0.405(5) | 1.33              | 2.68(1)   | 2.180(5)              | 2.170                 | 0.92              | 119(1)                       | 113.5                        |
| $Fe$-$N_3$ | 0.92(2)  | 1.33              | 2.41(2)   | 2.03(1)               | 2.014                 |                   |                              |                              |

Table III.- Octahedral geometry of the $Fe^{2+}$ environment: comparison between experiments and molecular dynamics.

| Distances (Å) | LS | | | HS | | |
|---|---|---|---|---|---|---|
| | $r^o$ | exp [a] | MD | $r^o$ | exp [b] | MD |
| $Fe-N_1$ | 2.105 | 1.965 | 1.967 | 2.586 | 2.251 | 2.250 |
| $Fe-N_2$ | 2.100 | 1.964 | 1.967 | 2.640 | 2.230 | 2.229 |
| $Fe-N_3$ | 2.067 | 1.938 | 1.934 | 2.288 | 2.040 | 2.041 |
| $N_1-N_1$ | | 2.695 | 2.695 | | 2.850 | 2.850 |
| $N_1-N_2$ | | 2.546 | 2.552 | | 2.707 | 2.707 |
| " | | 2.896 | 2.893 | | 3.246 | 3.246 |
| $N_1-N_3$ | | 2.831 | 2.828 | | 3.191 | 3.191 |
| $N_2-N_3$ | | 2.837 | 2.844 | | 3.030 | 3.030 |
| " | | 2.781 | 2.780 | | 3.293 | 3.293 |
| $N_3-N_3$ | | 2.695 | 2.694 | | 2.969 | 2.970 |
| Angles (deg) | | | | | | |
| $N_1-Fe-N_1$ | | 86.58 | 86.47 | | 78.58 | 78.60 |
| $N_1-Fe-N_2$ | | 80.79 | 80.86 | | 74.37 | 74.40 |
| " | | 94.98 | 94.61 | | 92.90 | 92.93 |
| $N_1-Fe-N_3$ | | 92.98 | 92.90 | | 95.97 | 95.96 |
| $N_2-Fe-N_3$ | | 90.88 | 90.89 | | 90.33 | 90.30 |
| " | | 93.27 | 93.54 | | 100.86 | 100.84 |
| $N_3-Fe-N_3$ | | 88.08 | 88.29 | | 93.36 | 93.35 |

[a] [21]
[b] [2]

Table IV.- *Fe-N* and *NCS* group frequencies (in cm$^{-1}$): comparison between experiments of Hoefer [30], molecular dynamics and *DFT*.

|  | *LS* | | | | *HS* | | | |
|---|---|---|---|---|---|---|---|---|
|  | exp | *MD* | *DFT* | | exp | *MD* | *DFT* | |
|  |  |  | I.R. | Raman |  |  | I.R. | Raman |
| *Fe-N* | 237 | 250 |  | 243 | 151 | 128 |  | 163 |
|  | 342 | 346 | 339 |  | 207 | 197 |  | 208 |
|  | 364 | 369 | 363 |  | 246, 259, 271 | 225 | 241, 257, 271 |  |
|  | 368 |  | 369 |  | 326 | 327, 334 |  | 323 |
| δ(N-C-S) | 447, 453, 458, 473, 487 | 418, 431, 446, 452 |  | 427, 435, 441 | 438, 469, 478 | 429, 460, 488 |  | 436, 442, 448 |
| *C-S* | 809, 831, 847 | 825, 845 |  | 770, 775 | 840 | 837, 846 |  | 779, 783 |
| *C-N* | 2124 | 2148, 2161 |  | 2098, 2109 | 2074, 2081 |  | 2093, 2111 | 2044, 2067 |

Table V.- Intermolecular interactions in the two spin states (For "chemical" clarity, the weak potential depths are given in kJ/mol units instead of eV).

|  | $\varepsilon°_{ij}$ (kJ/mol) | LS $\sigma_{ij}$ (Å) | HS $\sigma_{ij}$ (Å) |
|---|---|---|---|
| C...C | 0.393 | 3.040 | 3.420 |
| C...N | 0.357 | 3.043 | 3.570 |
| C...S | 0.752 | 3.040 | 3.741 |
| C...H | 0.208 | 2.900 | 2.900 |
| N...N | 0.324 | 3.220 | 3.662 |
| N...S | 0.685 | 3.400 | 3.846 |
| N...H | 0.208 | 2.900 | 2.913 |
| S...S | 1.439 | 4.030 | 4.030 |
| S...H | 0.234 | 2.470 | 2.900 |
| H...H | 0.038 | 2.150 | 2.150 |

Table VI.- Cell parameters and cell volume in *MD* for the *LS*-state (25 K and 140 K) and for the *HS*-state (140 K and 300 K); comparison with experiments.

|   | LS | | | HS | | |
|---|---|---|---|---|---|---|
|   | *MD* | exp | diff. (%) | *MD* | exp | diff. (%) |
| 25 K | | | | | | |
| $a$ (Å) | 12.1393(4) | 12.224(3) [a] | -0.7 | | | |
| $b$ (Å) | 15.5257(6) | 14.484(4) [a] | 7.2 | | | |
| $c$ (Å) | 17.4516(6) | 18.130(5) [a] | -3.7 | | | |
| $V$ (Å$^3$) | 3289.1(4) | 3210(1) [a] | 2.5 | | | |
| 140 K | | | | | | |
| $a$ (Å) | 12.287(1) | 12.370(3) [b] | -0.7 | 13.216(2) | | |
| $b$ (Å) | 16.610(2) | 14.764(3) [b] | 12.5 | 16.481(2) | | |
| $c$ (Å) | 16.146(2) | 18.281(4) [b] | -11.7 | 15.688(2) | | |
| $V$ (Å$^3$) | 3295(1) | 3339(2) [b] | -1.3 | 3417(1) | | |
| 298 K | | | | | | |
| $a$ (Å) | | | | 13.211(3) | 12.949(7) [b] | 2.0 |
| $b$ (Å) | | | | 15.818(3) | 15.183(2) [b] | 4.2 |
| $c$ (Å) | | | | 16.688(3) | 17.609(5) [b] | -5.2 |
| $V$ (Å$^3$) | | | | 3487(2) | 3462(2) [b] | 0.7 |

[a] [21]
[b] [2]

**Figure captions:**

Fig.1.- Molecular view of the *Fe(PM-BIA)$_2$NCS$_2$* complex. The numbering of the nitrogen atoms is the one found within the text, figures and tables.

Fig.2.- Fit of the quantum potential energy surface (*PES*) in the *LS* spin state by combining *Fe-N$_1$* Morse and Coulomb potentials with (straight red line in a) or without *N$_2$-Fe-C* harmonic potential (dotted blue line); by combining *Fe-N$_2$* Morse and Coulomb potentials with (straight red line in b) or without *N$_1$-Fe-C* harmonic potential (dotted blue line); by using *Fe-N$_3$* Morse and Coulomb potentials (straight red line in c). Black squares symbolize the *PES* values.

Fig.3.- Fit of the quantum potential energy surface (*PES*) in the *HS* spin state. Symbols have the same meaning as in Fig.2.

Fig.4.- Evolution of the *a* cell-parameter with temperature in the two spin states and comparison with diffraction data of [3] (a); evolution of the *b* cell-parameter with temperature in the two spin states (b); evolution of the *c* cell-parameter with temperature in the two spin states (c); evolution of the cell volume $V_{cell}$ with temperature in the two spin states (d). Symbol notations are: filled green circles: *LS*-field acting on *LS*-phase at 0 K with increasing temperature; filled blue triangles: *HS*-field acting on *HS*-phase with increasing temperature; empty symbols correspond to runs but in decreasing temperature from 298 K; filled red squares correspond to diffraction data.

Fig.5.- Evolution of the difference enthalpy between the two spin states versus temperature (heating runs: filled green circles; cooling-down runs: empty blue triangles). Comparison with *DSC* measurements (filled red squares [2, 31]).

Fig.6.- Evolution of the *a* cell-parameter with increasing pressure between 25K and 170 K ($LS_I$-field acting during MD runs) (a); evolution of the *b* cell-parameter (b); evolution of the *c* cell-parameter (c).

Fig.7.- Evolution of the *a* cell-parameter with increasing pressure between 190 K and 300 K ($HS_I$-field acting during MD runs) (a); evolution of the *b* cell-parameter (b); evolution of the *c* cell-parameter (c).

Fig.8.- (*P,T*) phase diagram showing four solid phases ($LS_I$: filled green circles; $HS_I$: filled blue triangles; $LS_{II}$: empty green circles; $HS_{II}$: empty blue triangles). Domains delimited by straight curves generate two triple points. Experimental points from [13, 24] are also shown for comparison ($LS_I$: filled red squares; $HS_I$: filled red down triangles; $LS_{II}$: empty red square; $HS_{II}$: empty red up triangles).

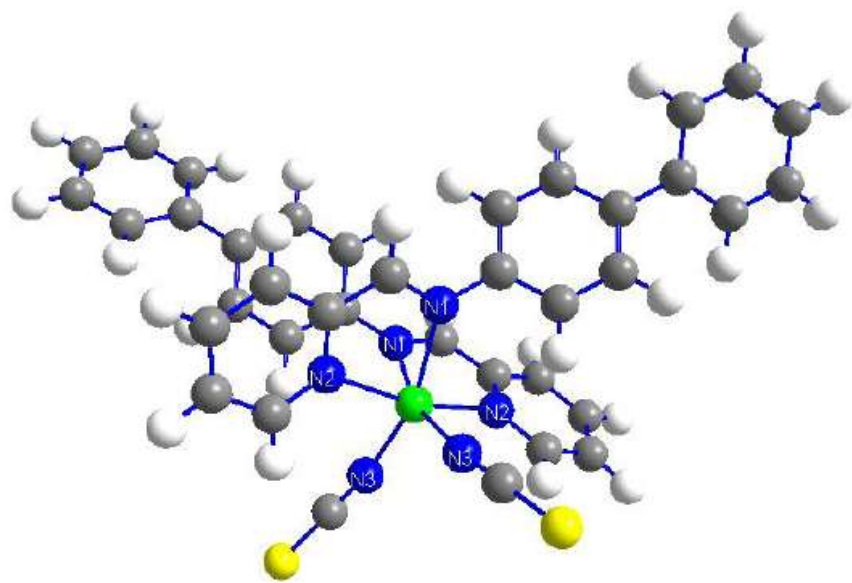

Fig.1

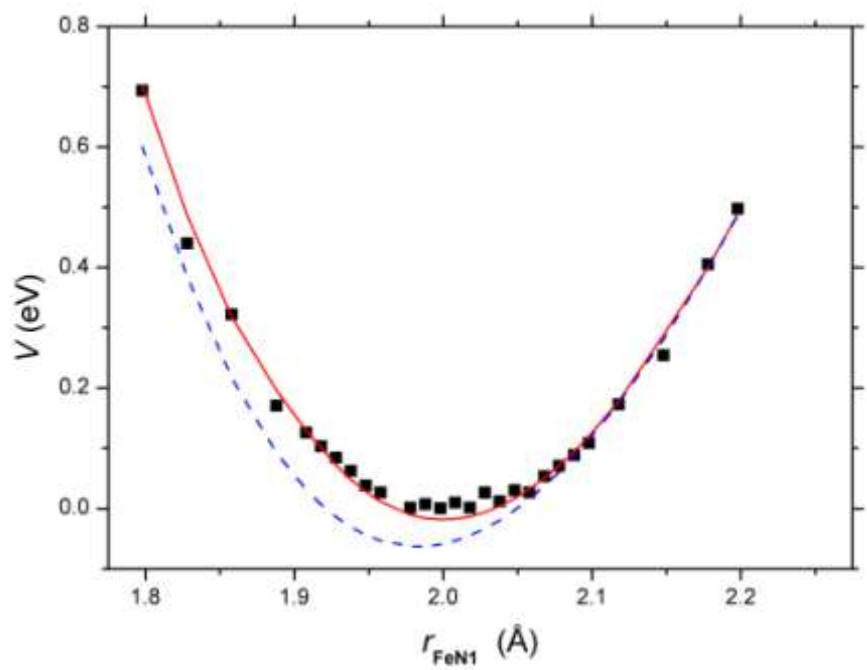

Fig.2a

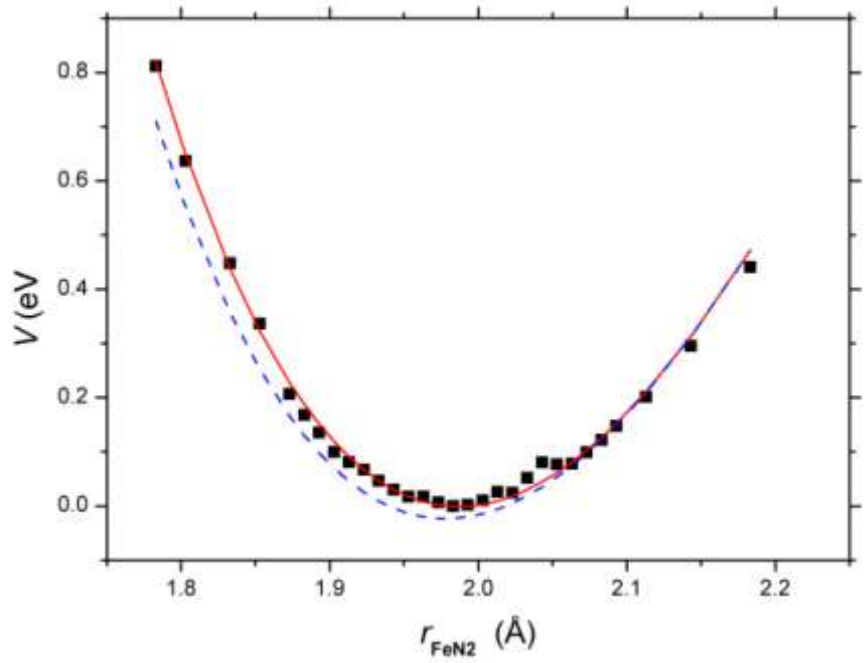

Fig.2b

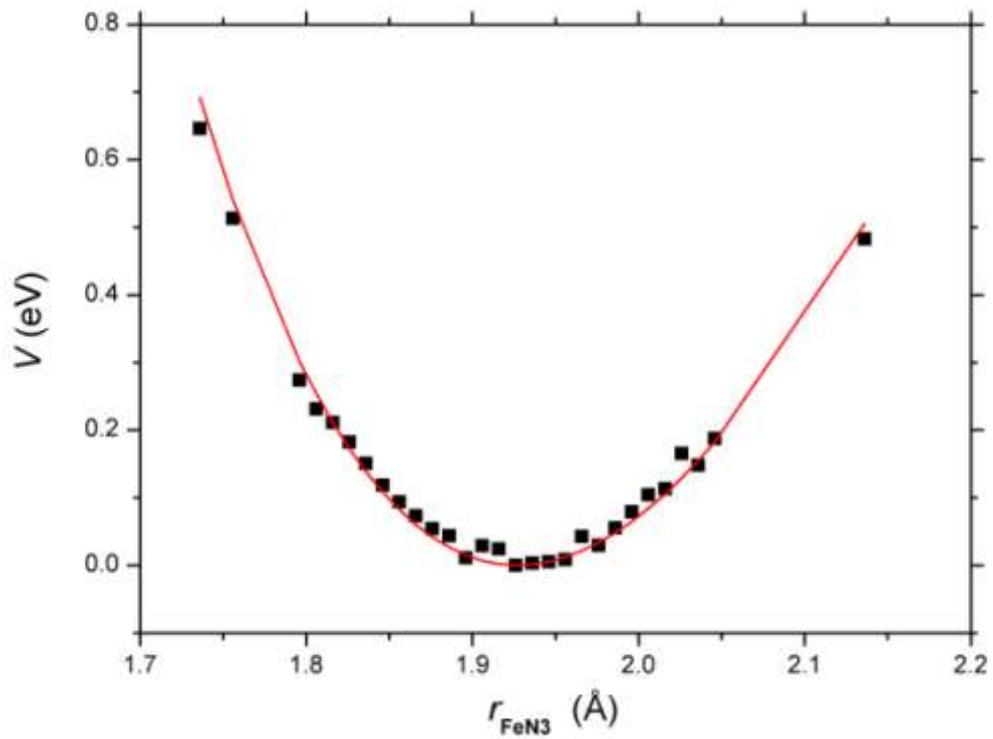

Fig.2c

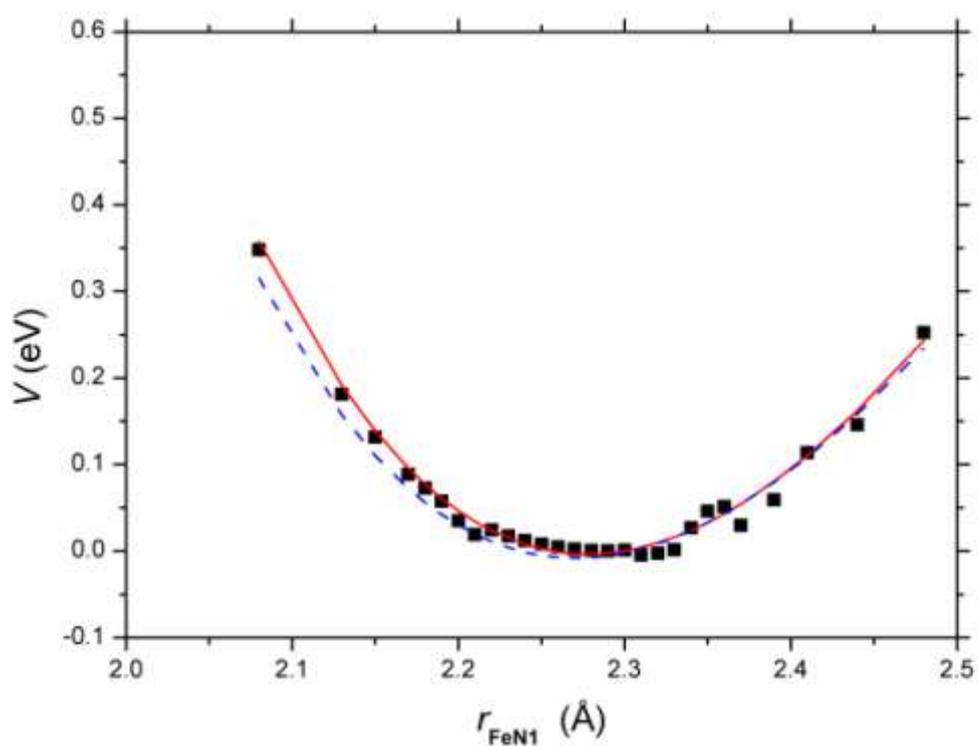

Fig.3a

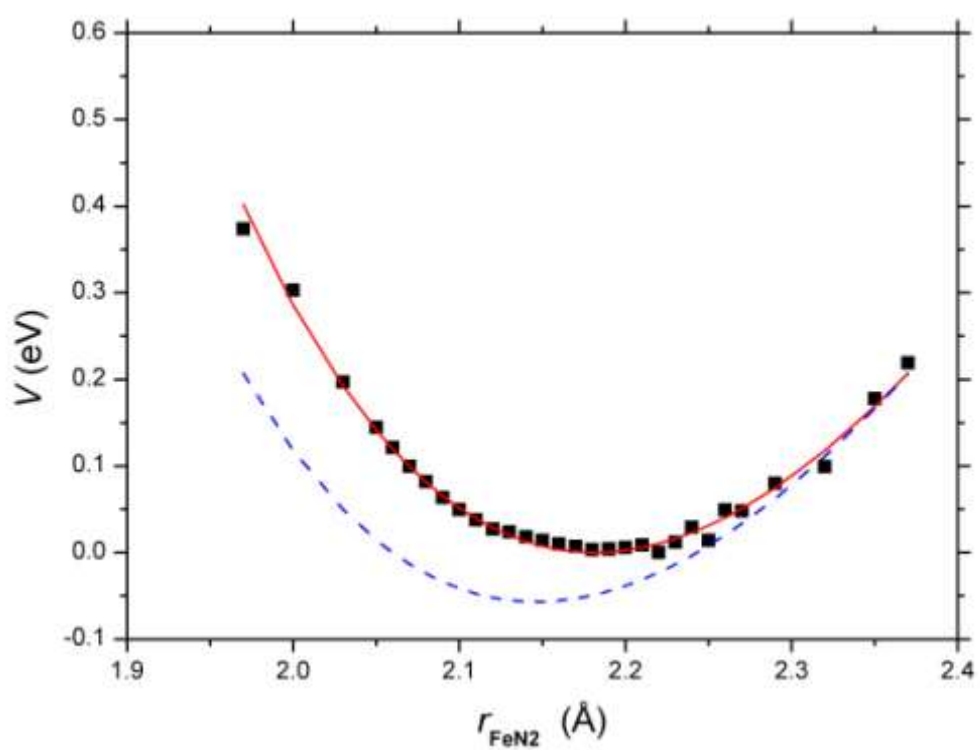

Fig.3b

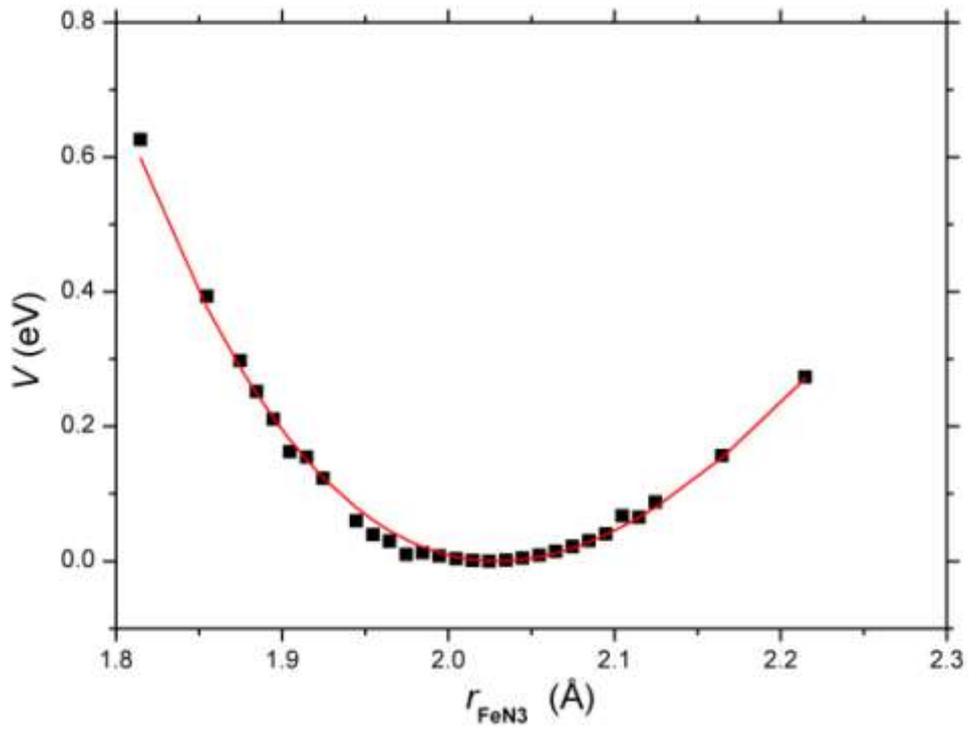

Fig.3c

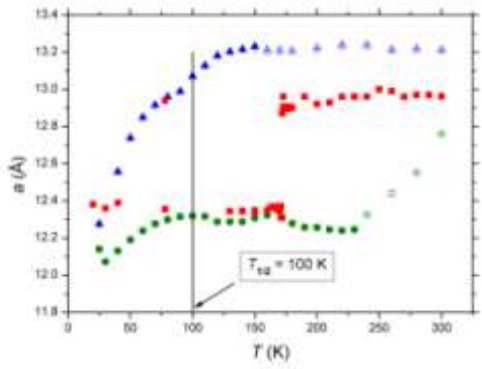

Fig.4a

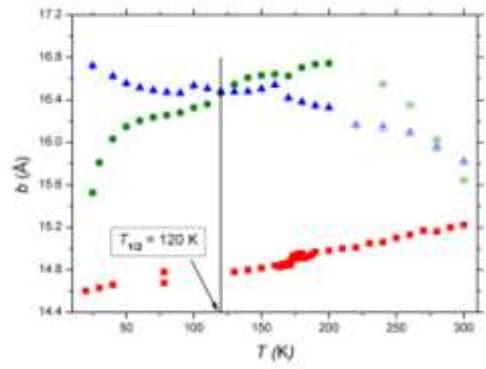

Fig.4b

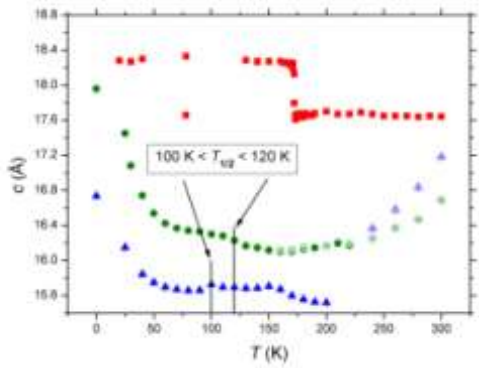

Fig.4c

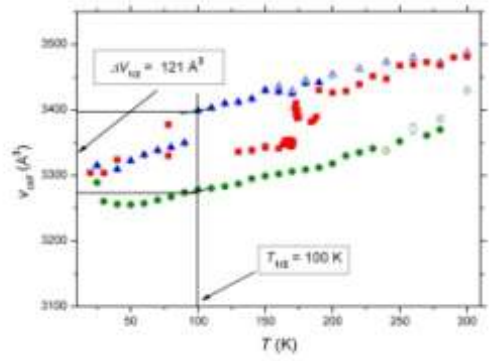

Fig.4d

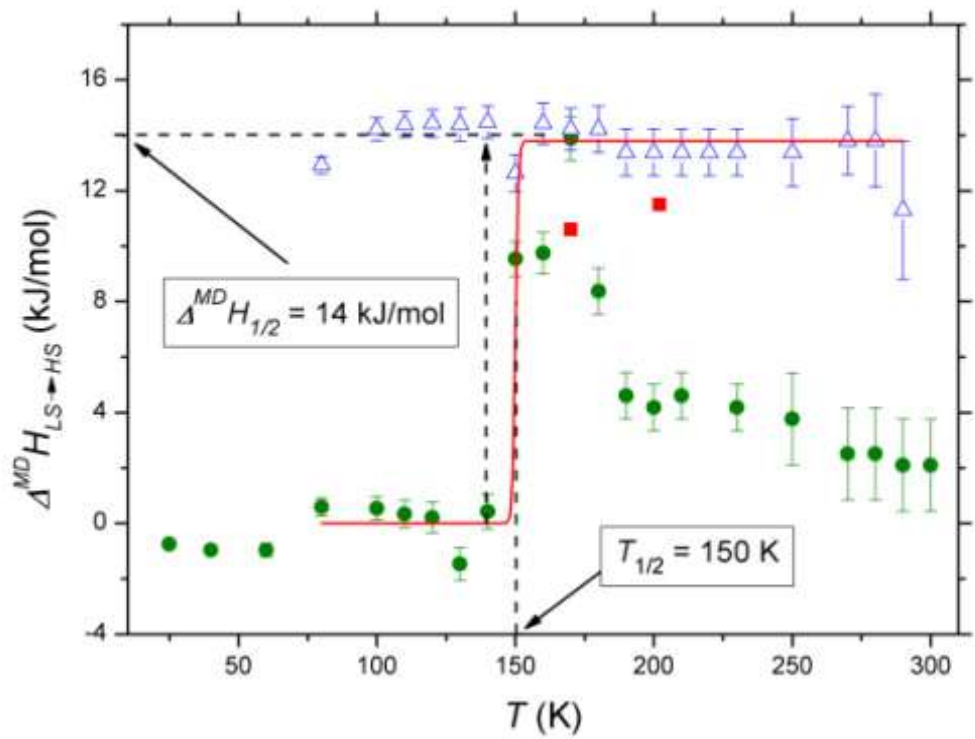

Fig.5

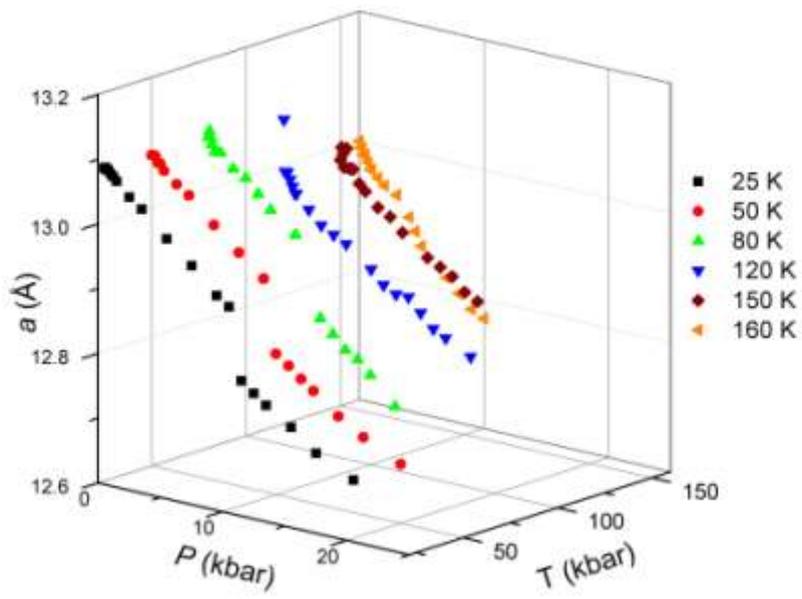

Fig.6a

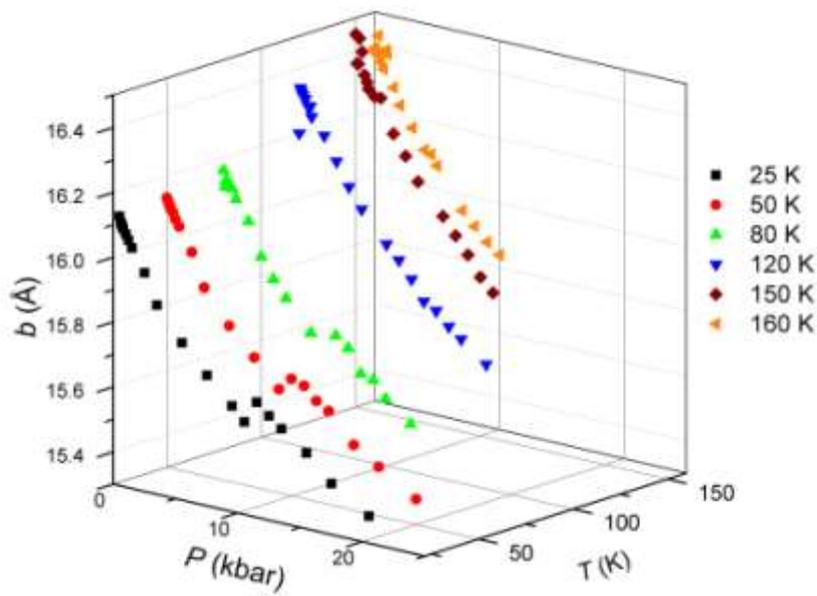

Fig.6b

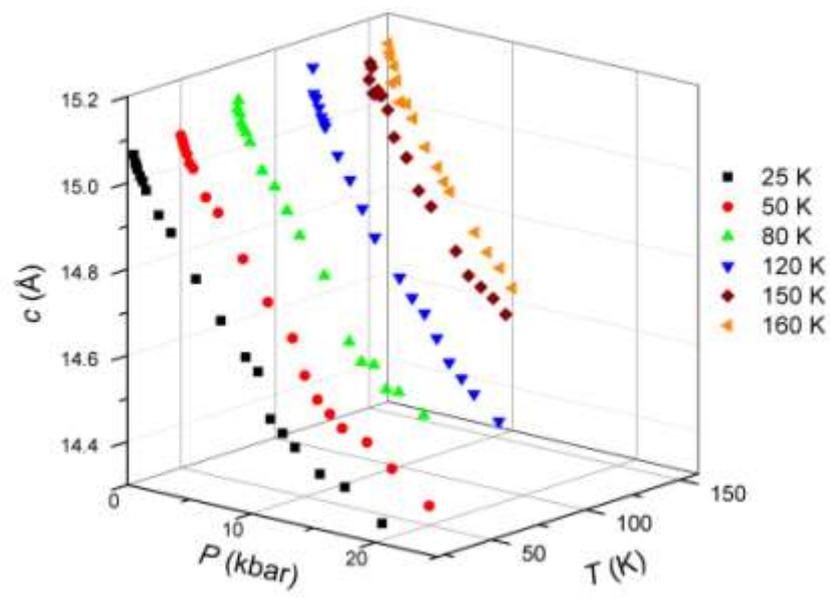

Fig.6c

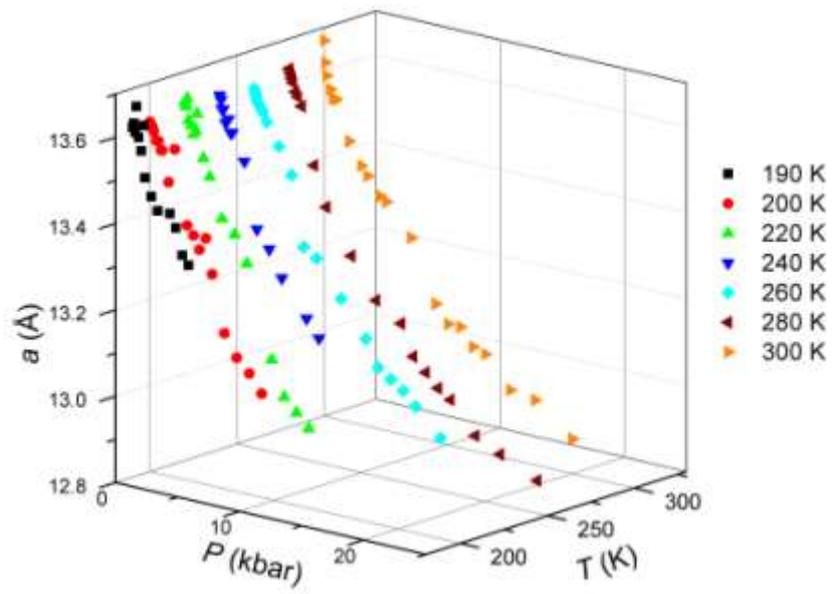

Fig.7a

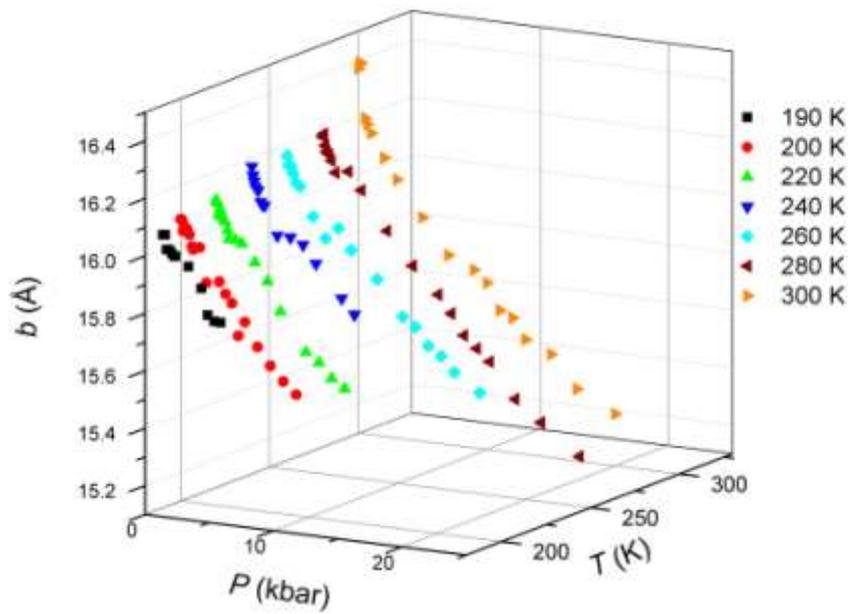

Fig.7b

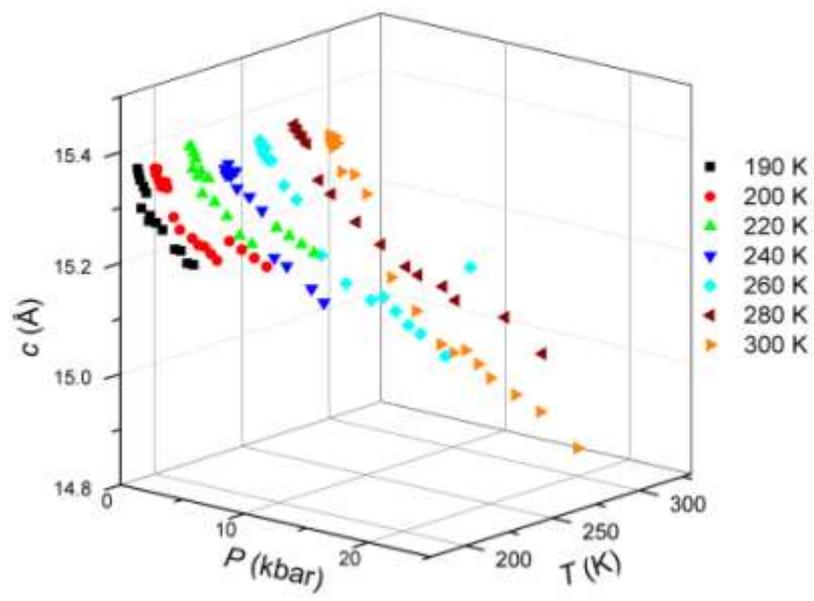

Fig.7c

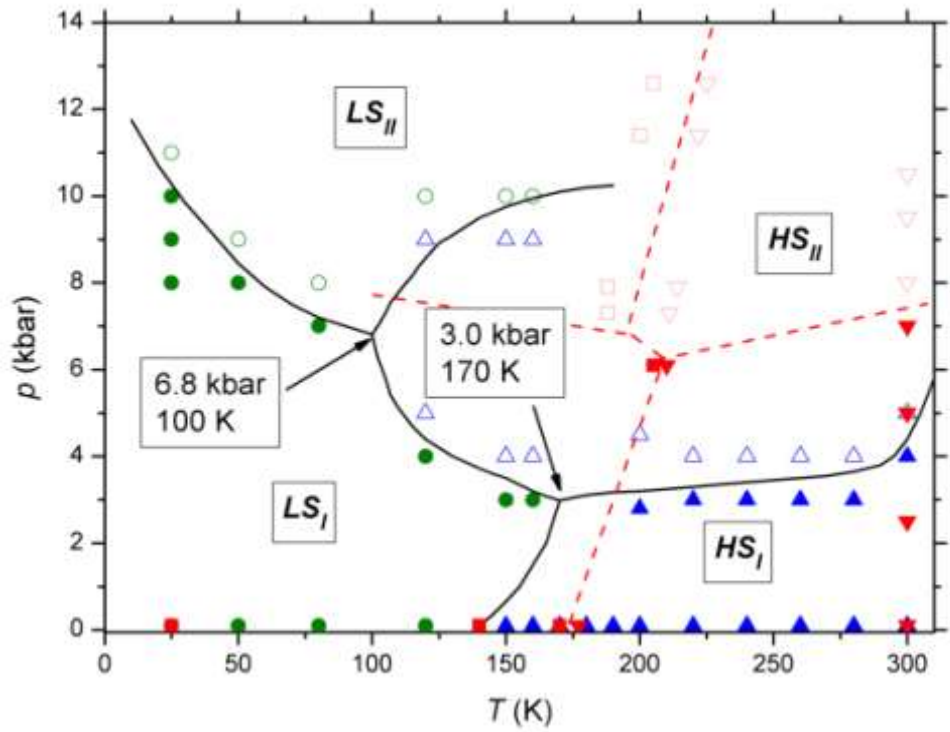

Fig.8

rr